\documentstyle[12pt,epsf]{article}
\newlength{\abstractwidth}
\newcommand{\OL}[1]{ \hspace{1pt}\overline{\hspace{-1pt}#1
   \hspace{-1pt}}\hspace{1pt} }

\renewcommand{\thefootnote}{\fnsymbol{footnote}}
\renewcommand{\thanks}[1]{\footnote{#1}} % Use this for footnotes
\newcommand{\starttext}{
\setcounter{footnote}{0}
\renewcommand{\thefootnote}{\arabic{footnote}}}
\renewcommand{\theequation}{\thesection.\arabic{equation}}
\newcommand{\be}{\begin{equation}}
\newcommand{\bea}{\begin{eqnarray}}
\newcommand{\eea}{\end{eqnarray}}
\newcommand{\beq}{\begin{equation}}
\newcommand{\ee}{\end{equation}}
\newcommand{\eeq}{\end{equation}}

\newcommand{\<}{\langle\,}

\renewcommand{\>}{\rangle}
\def\ba{\begin{eqnarray}}
\def\ea{\end{eqnarray}}

\def\14{{1\over4}}
\def\12{{1 \over 2}}
\def\eq{&=&}
\def\tro{\tilde{\rho}}
\def\d{\partial}

\def\h3{h^{3\over 2}}

\begin{document}
\renewcommand{\theequation}{\thesection.\arabic{equation}}
\begin{titlepage}
\bigskip
\rightline{NSF-ITP-01-185}
\rightline{SU-ITP 01/56}
\rightline{hep-th/0112204}

\bigskip\bigskip\bigskip\bigskip

\centerline{\Large \bf {String Theory and   }}
\centerline{\Large \bf {the Size of Hadrons }}

\bigskip\bigskip
\bigskip\bigskip
%\centerline{\it }
%\medskip
%\centerline{} \centerline{} \centerline{}
%\bigskip
\begin{center}
{\large Joseph Polchinski}\\ Institute for Theoretical
Physics, University of California\\ Santa Barbara, CA 93106-4030 \\ 
\vspace{.3 cm} {\large Leonard Susskind}\\ Department of Physics,
Stanford University\\ Stanford, CA 94305-4060 \\ \vspace{2cm}
\end{center}

\bigskip\bigskip
\begin{abstract}
We begin by outlining the ancient puzzle of off shell currents and
infinite size particles
in a string theory of hadrons. We then consider the problem from
the modern AdS/CFT perspective. We argue that although hadrons
should be thought of as  ideal thin strings from the 5-dimensional
bulk
point of view, the 4-dimensional strings are a superposition of
``fat" strings of different thickness.

We also find that the warped nature of the target geometry
provides a mechanism
for taming the infinite zero point fluctuations which apparently
produce a
divergent result for hadronic radii.

Finally a calculation of the large momentum behavior of the form
factor is given in the limit of strong 't Hooft parameter where
the classical gravity limit is appropriate. We find agreement with
parton model expectations.

\medskip
\noindent
\end{abstract}

\end{titlepage}
\starttext \baselineskip=17.63pt \setcounter{footnote}{0}

%%%%%%%%%%%%%%%%%%%%%%%%%%%%%%%%%%%%%%%%%%%%%%%%%%%%%%%%%%%%%%%%%%%%%%
%%%%%%
%%%%%%%%%%%%%%%%
%%%%%%%%%%%%%%%%%%%%%%%%%%%%%%%%%%%%%%%%%%%%%%%%%%%%%%%%%%%%%%%%%%%%%%
%%%%%%
%%%%%%%%%%%%%%%%
\setcounter{equation}{0}
\section{  The Puzzle of Infinite Size }

The historical origins of string theory were an attempt to
understand the structure of hadrons \cite{nambu} \cite{duality}.
However the theory
encountered a number of obstacles. Some obvious difficulties
involved the spectrum which invariably included massless vectors,
scalars and tensor particles. There were also more subtle problems
that seemed insurmountable and which involved the existence of
form factors. In other
words there was no possibility of constructing matrix elements of
local currents needed to describe the interaction of hadrons with
electromagnetism and gravitation \cite{holger}.
The natural candidates, vertex
operators like $\exp{ik X}$, can not be sensibly continued away
from specific discrete ``on shell" values of $k^2$. Closely connected
with this was the divergence encountered in attempting to compute
the hadronic electromagnetic or gravitational radius
\cite{holger} \cite{karliner}.
Thus string
theory was abandoned as a theory of hadrons and replaced by QCD.
The  success of string theory in understanding Regge trajectories
and quark confinement was understood in terms of an approximate
string-like behavior of chromo-electric flux tubes. According to
this view, hadronic strings are not the infinitely thin idealized
objects of mathematical string theory but are thick tubes similar
to the quantized flux lines in superconductors.
The ideal  string theory
was relegated to the world of quantum gravity.

However more recent developments \cite{maldacena} have strongly
suggested that an idealized form of string theory may exactly
describe certain gauge theories which are quite similar to
QCD \cite{confine,hard}.  We have returned
full circle to the suspicion that hadrons may be precisely
described by an idealized string theory, especially in the 't
Hooft limit \cite{tooft}. The new string theories are certainly more
complicated than the original versions and it seems very plausible
that the problems with the massless spectrum of particles will be
overcome. Less, however, has been studied about the problems
connected with local currents. In this paper we will show that the
new insights from the AdS/CFT correspondence suggest a solution to
the form factor problem.

We begin by reviewing the problem. For definiteness we work in the
light cone frame in which string theory has the form of a
conventional Galilean-invariant Hamiltonian quantum mechanics. The
degrees of freedom of the first-quantized string include
 $D-2$ transverse coordinates  $X^m(\sigma)$.  The
Lagrangian for these variables is
\be
L={1\over 4 \pi}\int_0^{2\pi P_-} \! d\sigma\, \Bigl(\dot{X}\dot{X}-
\alpha'^{-2}X'X' \Bigr)\ ,
\ee
where $\dot{X}$
and $X'$ denote derivatives with respect to light-cone time $\tau$
and string parameter $\sigma$. The light-cone momentum $P_-$ is
conjugate to the light-like coordinate $x^-$. All irrelevant constants
have been
set to unity.

An important feature of the light-cone theory involves the local
distribution of $P_-$ on the string. The rule is that the
distribution of $P_-$ is uniform with respect to $\sigma$. In
other words the longitudinal momentum $dP_-$ carried on a segment
of string $d \sigma $ is exactly $d \sigma /{2 \pi}$.
Let us now consider the transverse density of $P_-$. In a space-time
field theory this would be given by
\be
\rho(X) = \int dx^-\, T_{--}(X,x^-)\ ,
\ee
where $T$ is the energy momentum tensor of the field theory, and $X$
without a superscript denotes the two noncompact transverse
coordinates. Matrix elements of $\rho$ between states of equal $P_-$
define form factors for gravitational interactions of the state and
are entirely analogous to electromagnetic form factors.

The natural object in string theory to identify with $\rho(X)$
is
\be
{1\over 2\pi}\int_0^{2\pi P_-} \! d\sigma \, \delta^2
(X-X(\sigma)) \ .\label{rho}
\ee
(To be precise, we subtract the center-of-mass mode from
$X(\sigma)$.) In other words  $\rho(X)$ receives contributions from
every element of string localized at $X$. The Fourier transform of
$\rho(X)$
\be
\tro(k)= {1\over 2\pi}\int_0^{2\pi P_-} d \sigma \exp{i k X(\sigma)}
\label{tro}
\ee
defines a system of form factors by its matrix elements between
string states.

The mean square radius of the distribution function is given by
\be
\OL{r^2}= \int d^{2}X \, X^2 \< \rho(X) \>
\ee
and can be rewritten in terms of $\tro$,
\be
\OL{r^2} = -\partial_k \partial_k \<\tro(k) \> |_{k=0}\ .
\label{radsq}
\ee
Eq.~(\ref{radsq}) is the standard definition of the mean-square radius
in terms of the momentum space form factor.
The squared radius is also given by
\be
\< X(\sigma)^2 \>\ .
\ee
where the value of $\sigma$ is arbitrary. 

For a field theory with a mass gap, such as pure QCD, it is
possible to prove that $\OL{r^2}$ is finite. This follows from the
standard analytic properties of form factors.
The problem arises when we attempt to apply the world sheet field
theory to compute $\< X(\sigma)^2 \>$. An elementary calculation
based on the oscillator representation of $X$ gives a sum over
modes
\be
\<X^2\> \sim  \alpha'\sum_0^{\infty} {1 \over n} =\alpha'
\log{\infty}\ .
\ee
A related disaster occurs when we compute the form factor which is
easily seen to have to form
\be
\< \tro(k) \> = \exp{(-k^2 \<X^2 \>/2)}\ .
\ee
Evidently it is only non-zero  at
$k^2=0$.

In a covariant description of string theory the problem has its
roots in the fact that the graviton vertex operator is
only well defined on
the mass shell of the graviton, $k^2=0$. Vertex operators to be
well defined must correspond to perturbations with vanishing world
sheet $\beta$-function. This implies that they should
correspond
to on shell solutions of the appropriate space-time gravitational
theory. For the kinematical situation in which the graviton
carries vanishing $k_{\pm}$ the transverse momentum must vanish.
Thus, no well defined off shell continuation of the form factor
exists.

One might wonder if the divergence of $X^2$  is special to the case of
a free world sheet field theory. The answer is that the divergence
can only be made worse by interactions. The 2-point function of a
unitary quantum field theory is at least as divergent as the
corresponding free field theory. This follows from the spectral
representation for the two point function and the positivity of
the spectral function. Thus it is hard to see how an ideal string
theory can ever describe hadrons.

\setcounter{equation}{0}
\section{ Light Cone Strings in AdS }

There are good reasons to believe that certain confining
deformations of maximally supersymmetric Yang Mills theory are
string theories albeit in higher dimensions. The strings move in a
five-dimensional space
that is asymptotically AdS.\footnote{
Strictly speaking the target space is
ten-dimensional with the form $AdS_5$ times a compact space such as
$S_5$. In this paper the compact factor plays no role.}
In the 't Hooft
limit these theories are believed to be free string theories.
Evidently if this is so there must exist a well defined string
prescription for form factors in the 4-D theory.

What we will see is that although the theory in bulk of AdS is an
ideal thin-string theory, the 4-D boundary field theory is not
described by thin strings. That may seem surprising. Suppose that
in the light-cone frame the thin 5-D string has the form
\be
(X(\sigma),\, Y(\sigma))
\ee
where $X$ are the transverse coordinates of 4-D Minkowski space
and $Y$ is the additional coordinate perpendicular to the boundary
of AdS. Then it would seem natural to consider the projection of
the string onto the $X$ plane to define a thin string. According to
this
view the mean-squared radius would again be $\<X^2\>$ and we would
be no better off than before. Before discussing the resolution of
this problem let us work out the bosonic part of the light-cone string
Lagrangian in
AdS. We will make no attempt to derive the full supersymmetric
form of the theory \cite{thorn} in this paper. We believe the resolution
of the form factor problem does not require this. On the geometric side
we will also ignore the 5-sphere component of the geometry implied
by the usual R-symmetry of the $N=4$ supersymmetry.

The metric of AdS is given by
\be
ds^2 = R^2 {2 dx^+ dx^- +dX^2 +dY^2 \over Y^2}\ .
\ee
We have defined the overall scale of the AdS (radius of curvature)
to be $R$.

In order to pass to the light cone frame we must also introduce
the world sheet metric $h_{ij}$. In the usual flat space theory it
is possible to fix the world sheet metric to be in both the light
cone gauge $\sigma_0 = \tau = x^+$ and also the conformal gauge
$h_{00} = -h_{11}, \ h_{01}=0$. However this is not generally
possible since it entails three gauge conditions which is one too
many (we do not count the local Weyl symmetry, because none of these
conditions fix it; this would require a fourth condition, $\det h = -1$).
The special feature of flat space which permit the over-fixing of the
gauge is not shared by AdS. Thus we must give up the conformal gauge if
we wish to work in light-cone gauge.

Let us fix the gauge by choosing two conditions
\bea
\sigma_0 \eq x^+\ , \cr
h_{01} \eq 0\ . \label{gauge}
\eea
The Lagrangian becomes
\be
L={1\over 4 \pi}\int d\sigma\, \biggl\{ \frac{R^2 E}{Y^2} (2\dot X^- +
\dot{X}\dot{X} + \dot Y \dot Y) - \frac{1}{\alpha'^2 R^2 EY^2}
(X'X' + Y' Y') \biggr\}\ ,
\ee
where we have defined
\be
\sqrt{-{h_{11}\over h_{00}}} =E\ .
\ee
The equation of motion from $X^-$ states that the density $p_-(\sigma)
= R^2 E(\sigma)/2\pi Y^2(\sigma)$ is time-independent.  We can then
use the
$\tau$-independent $\sigma$ reparameterization not fixed by
eq.~(\ref{gauge}) to set $p^- = 1/2\pi$.  The Hamiltonian becomes
\be
H=
\frac{1}{2} \int_0^{2\pi P_-} \! d\sigma \,\biggl\{ P_X P_X +P_Y P_Y
+{R^4\over \alpha'^2 Y^4} ( X' X' + Y' Y') \biggr\}\ .
\ee
This is a more or less conventional string action with the unusual
feature that the effective string tension scales like $1/Y^2$.
Thus the tension blows up at the AdS boundary $Y=0$ and tends to
zero at the horizon $Y= \infty$. This of course is a manifestation
of the usual UV/IR connection.

The  Hamiltonian could be obtained from an action
\be
S= \frac{1}{2} \int_0^{2\pi P_-} \! d \sigma d\tau \biggl\{
\dot{X}\dot{X} +\dot{Y}\dot{Y} -{R^4\over \alpha'^2  Y^4}
(X' X' + Y' Y' ) \biggr\}\ . \label{gfact}
\ee
This action thought of as a $1+1$ dimensional field theory is not
Lorentz invariant in the world-sheet sense. However it is
classically scale invariant if we assume $X$ and $Y$ are dimension
zero. The Hamiltonian has dimension $1$ and therefore scales as
the inverse length of the $\sigma$ circle which is  $\sim P_-$.
We recognize this scale symmetry as spacetime longitudinal boost
invariance
under which $H$ and $P_-$ scale oppositely and $X,Y$ are
invariant. No doubt the actual
Lagrangian, properly supersymmetrized, retains this symmetry
when quantized.

Let us consider the equal time correlation function
$\<X(0)X(\sigma)\>$
in the field theory defined by eq.~(\ref{gfact}), or more precisely in
its supersymmetrized version. By inserting a complete set of
eigenstates of the energy and (world sheet) momentum we obtain
\bea
\<X(0)X(\sigma)\>&=& \sum \int {dE\, dp\over p^2}
e^{ip\sigma } |\<X(0)|E,p\>|^2  \cr
&=& \int {dE\, dp\over p^2}e^{ip\sigma } F(E,p)
\label{spect}
\eea
with $F\geq 0$.  The measure of integration $dE dp/p^2$ follows from
the fact that $X$ has ``engineering" dimension zero under the
longitudinal boost rescaling. Furthermore the assumption that the
scale invariance is preserved in the quantum theory requires $F(E,p)
= F(E/p)$ for large $p,E$. It follows that as long as $F$ does not go
to zero in this limit that the correlation function diverges as
$\sigma \to 0$. This would imply $X^2 = \infty$.

\setcounter{equation}{0}
\section{Dressing the Vertex with $Y$  Dependence }

Let us consider the problem from the point of view of the vertex
operator $\exp{ikX}$. One problem that we have emphasized is that it
is not a solution of the on-shell condition. We can try to fix
this by replacing it with a solution of the wave equation for a
graviton in AdS space~\cite{Callan}. The relevant equation is
\be
\left(
\d_{\mu}\d^{\mu} +
Y^3 \d_Y Y^{-3}\d_Y
\right)\Phi =0\ ,
\ee
where $\mu$ runs over the four dimensions of flat Minkowski space.
To be precise, this is the first point at which we have assumed large
't Hooft parameter, so that we can use the low energy field equations.

The particular solutions we are looking for are independent of the
$x^{\pm}$ and have the form
\be
\Phi = \exp{(ikX)} f(k,Y)\ ,
\ee
where $F$ satisfies
\be
Y^3 \d_Y Y^{-3} \d_Y f(k,Y) = k^2 f(k,Y)\ . \label{minsca}
\ee
Thus the on shell vertex operator has the form
\be
\int d\sigma \exp{(ikX(\sigma))} f(k,Y(\sigma))\ .
\ee
The factor $f$ is a  dressing of the vertex, necessary to
make its matrix elements well defined for $k \neq 0$.

Let us consider the mean square radius of the hadron defined by
eq.~(\ref{radsq}),
\bea
\OL{r^2}&=& -\d_k \d_k \<  f(k,Y) \exp{(ikX)}\> |_{k=0} \nonumber\\
&=&\< X^2 f(0,Y) -2iX\cdot f'(0,Y) -f''(0,Y) \>
\eea
where $f''\equiv \d_k\d_k f$.

For a state of zero angular momentum in the $X$ plane the term
linear in $X$ vanishes and we have
\be
\OL{r^2}=\< X^2 f(0,Y) -f''(0,Y) \>.
\ee
To compute $f$ and $f''$ we Taylor expand $f(k,Y)$ in powers of
$k$. There are two linearly independent
solutions,
\be
f(k,Y) = Y^4 + {1\over 12}k^2 Y^6 +\cdots
\ee
and
\be
f(k,Y) =1 - {1\over 4}k^2 Y^2 +\cdots\ .
\ee
Only the second of these is relevant to the problem of form factors.
To see this we need only note that the vertex at $k=0$
is just the operator that measures $P_-$. For states with
$P_-=1$ this operator in just the identity. This implies that
$F(0,Y) =1$.  Thus we find
\be
\OL{r^2}=\<\, (X^2  +Y^2) \, \>\ . \label{dressed}
\ee

This dressed result should be finite, because the
on-shell condition in five (actually ten) dimensions arises from the
requirement of world-sheet conformal invariance.  One might have
expected that this would happen as a simple cancellation between a
divergence in $X^2$ and a divergence in the dressing.  However, this
is not the case: $\OL{r^2}$ as given in eq.~(\ref{dressed}) is the
sum of
$X^2$ and a positive term.  Evidently cancellation is
not the answer, and it must be that, in contrast to all experience in
field theory, the operator $X^2$ is finite!

The dressing of the vertex by the factor $f(k,Y)$ obviously
modifies the expression~(\ref{rho})
for the transverse density $\rho$. If we define the Fourier
transform of $f$ with respect to $k$ to be $\tilde{f}(X,Y)$,
eq.~(\ref{rho}) is replaced by
\be
\rho(X) \sim \int_0^{2\pi P_-} \!  d \sigma\,
\tilde{f}(X-X(\sigma),Y).
\ee
This means that an ideal thin string in the AdS bulk space is
smeared out by the holographic projection onto the boundary. This
is of course the familiar UV/IR correspondence at work. Bulk
strings near the boundary are projected as very thin strings in
the 4-D theory but those far from the boundary are fat. The extra
term $\<Y^2\>$ in eq.~(\ref{dressed}) represents this fattening.
Evidently we have only made things worse by including the dressing.

Before discussing the solution to the problem let us make some
remarks about confining deformations in the context of AdS/CFT.
Bulk descriptions of  confining deformations of super Yang Mills theory
have an effective infrared ``wall" at
a value of $Y$ which represents the confinement  scale. In these
cases
the metric (2.2) is modified in the infrared region.  A simple
model for this is
\be
ds^2= h(Y) \left( 2 dx^+dx^-  + dX^2 + dY^2 \right)
\ee
where, as in the conformal case, $h\sim 1/Y^2$ for $Y \to 0$.
Assume that $h$ has a minimum at the confinement scale, $Y=Y^*$.

The light-cone hamiltonian is easily worked out,
\be
H= \frac{1}{2} \int_0^{2\pi P_-} \!  d\sigma
\left\{ P_X P_X +P_Y P_Y +h(Y)^2 \alpha'^{-2}
(X' X' + Y' Y') \right\}
\ee
Consider a string stretched along the $X$ direction and choose $\sigma$
so that $\d_{\sigma}X=1$. The potential energy of the string is
then given by
\be
V(Y)=h(Y)^2 \alpha'^{-2}
\ee
which has a minimum at $Y=Y^*$. Thus  a classical long
straight string will be in equilibrium at this value of $Y$.
This classical bulk string
corresponds to a field theory configuration which, according to
the UV/IR connection, is thickened to a size $\sim Y^*$, that is,
the QCD scale.

Quantum fluctuations will cause the wave
function of the string to fluctuate away from $Y^*$.
The implication is that the
QCD string is a superposition of different thickness values
extending from infinitely thin to QCD scale. Indeed different
parts of the string can fluctuate in thickness over this range.
The portions of the string near $Y=0$ will be very thin and will
determine the large momentum behavior of the form factor.

\setcounter{equation}{0}
\section{Finiteness of $\< X^2 \>$}

We believe that despite the argument given at the end of Section 2
the value of $\< X^2 \> $ is finite. This can only be if the
function $F=\sum |\<X(0)|E,p\>|^2$ vanishes in the scaling limit
of large $E,p$. We will first give an intuitive argument and
follow it with a more technical renormalization group analysis.

First suppose the string is ``stuck" at some value of $Y$. In that
case the action  for $X$ in eq.~(\ref{gfact}) is a conventional string
action except that the sting tension is replaced by $1/Y^4$. The
divergence in $X^2$ would then be given by
\be
\<X^2 \> = Y^2 |\log{\epsilon}|.
\ee
If we ignore quantum fluctuations of $Y $ we could replace $Y$ by
$Y^*$. But $Y$ fluctuates as well as $X$ and can be expected to
fluctuate toward the boundary as $\epsilon$
tends to zero. This is just the usual UV/IR connection in AdS.
Therefore as we remove the cutoff the fluctuations of $X$ are
diminished because the string moves into a region of increasing
effective
stiffness. If for example the average value of $Y^2$ tends to zero
as $|1/\log{\epsilon}|$ or faster then the fluctuations of $X$ would
remain
bounded.  

To see that this happens we consider the renormalization
running of the operator $X^2$.
We begin with the bare theory defined with a cutoff length $\epsilon$
on the world sheet. We can then ask how a given operator in this
bare theory is described in a renormalized version of the theory
with a cutoff at some longer distance $l$. A general operator
$\phi(X,Y)$ runs to lower momentum scales
according to the renormalization group equation
\be
(l \partial/\partial l) \phi(X,Y,l) =
(\alpha'/2) \nabla^2 \phi(X,Y,l)\ .  \label{rng}
\ee
For example, consider flat space and the operator $X^2$. We look
for a solution of~(\ref{rng}) with
\be
\phi(X, \epsilon) =X^2\ .
\ee
The solution is
\be
\phi(X, l) = X^2 +\alpha' \log{l/{\epsilon}}.
\ee
Thus if we regulate the theory at some fixed scale,
for example $l\sim 1$, the matrix elements of $X^2$ blow up as we
send $\epsilon \to 0$.

By contrast, consider the the case of AdS space where
\be
\nabla^2 = R^{-2} ( Y^2 \partial_X^2 + Y^5 \partial_Y Y^{-3}
\partial_Y)\ .
\ee
For a solution of the form $X^2 + f(l) Y^2$ this becomes
\be
(l \partial/\partial l) f = (2\alpha'/R^2) (1 - f)\ .
\ee
With $f(\epsilon) = 0$ the solution is
\be
 f(l) =  1 - (\epsilon/l)^{2\alpha'/R^2}\ .
\ee
So if we fix the scale $l$ and take the cutoff length $\epsilon$
to zero the matrix elements tend to finite
limits and the problem of infinite
radii is resolved.  If, however,
we expand in powers of
$\alpha'$ there are logarithmic divergences.
Note that the operator $X^2$ runs to a fixed point $X^2+Y^2$ which
is just the operator in eq(3.10) which represents the mean squared
radius $\OL{r^2}$.\footnote{Juan Maldacena raised the issue as to
whether the undressed operator $X^2$ is physically observable.
Once light-cone gauge is fixed, every operator should be observable.
A world-sheet scale transformation corresponds to a spacetime
boost, so a scale-dependent operator has a nontrivial
transformation under this boost.}

Let us make a similar renormalization group calculation
of the two-point function.  For simplicity we consider the
equal-time case; the method extends to unequal times, but the
expressions are not as simple.  When the renormalization scale is
the same as the separation there are no large logarithms and so we
obtain the semiclassical result
\be
{X^i_\sigma}'(\sigma,0) X^j_\sigma(0,0) \sim - \delta^{ij}\frac{\alpha'
Y_\sigma^2}{\sigma R^2}
\ .
\ee
Subscripts on operators signify the renormalization scale.  
The $\sigma$-derivative (prime) is taken in order to make this
insensitive to IR physics.  Now let
us write this in terms of operators renormalized at a lower scale
$l$ as in eq.~(\ref{rng}).  The operator $X$ renormalizes
trivially, and so we can drop the subscript.  The renormalization of $Y^2$
gives
\be
{X^i}'(\sigma,0) X^j(0,0) \sim - \delta^{ij}
(\sigma/l)^{2\alpha'/R^2}
\frac{\alpha' Y_l^2}{\sigma R^2}\ . 
\ee
We can now integrate with respect to $\sigma$,
\be
X^i(\sigma,0) X^j(0,0) \sim 
{\rm constant} - \frac{\delta^{ij}}{2} (\sigma/l)^{2\alpha'/R^2}
{Y_l^2} \ .  \label{twopoint}
\ee
To complete the evaluation of the two-point function we must run
the operators down to the length scale of the string, and then
evaluate the expectation value of $Y^2$ in
the given string state, which we might take to be a wavepacket
centered on some value of $Y$,
\be
\< X^i(\sigma,0) X^j(0,0) \> \sim 
{\rm constant} - \frac{\delta^{ij}}{2} \sigma^{2\alpha'/R^2}
\< Y^2 \> \ .  \label{twopointev}
\ee
The result has the
advertised property: a finite limit as $\sigma \to 0$.

Precisely where does the spectral argument~(\ref{spect}) go wrong?  It is
in the assumption of scale invariance.  Scale invariance is of course
broken by the finite coordinate length of the string, but this is
an IR effect that should not change the short-distance
behavior of correlators.  In the present case there is a second
source of scale breaking.  The operators $Y^q$ have anomalous
dimension $(2\alpha'/R^2) q (4-q)$.  Since $Y$ is positive, these
operators must have expectation values in any state, and this
breaks the scale invariance spontaneously.  We see this in
eq.~(\ref{twopoint}): this equation is covariant under scale
transformations, but when the operators on the right are replaced
with their expectation values the two-point function scales
nontrivially.  (One might try to interpret the
result~(\ref{twopointev}) in terms of an anomalous dimension for $X$,
but this does not seem sensible).
Note the close analogy between $Y$ and a Liouville
field, as emphasized in ref.~\cite{GKP}.

The reader may wonder how the finiteness of $X^2$ can be explained
in covariant gauges such as the conformal gauge in which the world
sheet theory has the form of a relativistic field theory. A
standard argument
insures that the singularity in a two point function can not be
less singular than a free field; in this case logarithmic.
The argument is based on the positivity of spectral functions
which in turn assumes the metric in the space of states is
positive. In general this is not the case in covariant gauges.

\setcounter{equation}{0}
\section{Calculation of the Form Factor}

At large 't Hooft parameter, we can use the dual supergravity
description to calculate the form factor.  To calculate a matrix
element
\be
\< A | T_{--}(q) | B \>
\ee
we use the prescription of ref.~\cite{GKP,wit}.  That is, the
hadronic states $A$ and $B$ correspond to normalizable string
states in the bulk, while the local operator insertion corresponds
to a modification of the boundary conditions such that the
nonnormalizable modes are excited.  In particular, a perturbation
of the bulk metric to
\be
ds^2 = R^2 {2 dx^+ dx^- + 2 f(X,Y) dx^+ dx^+
+dX^2 +dY^2 \over Y^2}
\ee
corresponds to a perturbation of the boundary metric,
\be
ds^2_{\rm boundary} = 2 dx^+ dx^- + 2 f(X,Y) dx^+ dx^+ dX^2 \ .
\ee
This in turn corresponds to an insertion of the operator
\be
- i \int dx^+ dx^- d^2 X\, f(X,0) T_{--}(x^+,x^-,X)\ .
\ee

The metric perturbation satisfies the same minimal scalar
equation~(\ref{minsca}), as a consequence of the field equation
${\cal R}_{++} = 0$.  In
$AdS_5$ the nonnormalizable solution would be
\be
f(X,Y) = \int e^{iqX} q^2 Y^2 K_2( qY)\ .
\ee
In nonconformal theories this is modified at $Y
\stackrel{\sim}{>} Y^*$.  Let us first consider hard scattering,
where $q Y^* \gg 1$.  The metric perturbation is exponentially
suppressed in $q Y^*$ when $Y \stackrel{\sim}{>} Y^*$, and so
we can calculate as though in $AdS_5$ (there is an admixture of the
normalizable mode, but again this is exponentially suppressed).

A canonically normalized bulk scalar field $\Phi$ couples
to the metric perturbation as
\be
- i \int dx^+ dx^- d^2 X dY\,\frac{R^3}{Y^3} f(X,Y) \partial_- \Phi
\partial_- \Phi\ . \label{olap}
\ee
A gauge theory hadron corresponds to a superposition of such states,
with the mixing being due to the conformal symmetry breaking.  For
hard scattering, $f(X,Y)$ cuts off the integrand at $Y > q^{-1}$ and
so we are interested in the contribution that is least suppressed at
small $Y$.  As in ref.~\cite{hard}, if we restrict to scalar
fields then the dominant component is that of smallest conformal
dimension $\Delta$.  At small $Y$ the scalar wavefunction is
\be
\Phi \propto R^{-3/2} Y^* (Y/Y^*)^\Delta \exp(i p_\mu x^\mu)\ .
\label{tails}
\ee
The normalization is determined as in ref.~\cite{hard}, though it is
written differently here because we have reduced to five dimensions
from the start.

The overlap integral~(\ref{olap}) is peaked at $Y \sim q^{-1}$, giving
\be
\< A | T_{--}(q) | B \> \propto p_-^2 (q^2 Y^{*2})^{1-\Delta}\ .
\ee
The numerical normalization depends on the coefficient of the
leading piece~(\ref{tails}) of the wavefunction, which is
determined by the details of the geometry at $Y \stackrel{\sim}{>}
Y^*$. As with exclusive scattering \cite{hard}, the behavior agrees
with QCD, with the dimension $\Delta$ interpreted as the number of
partons. Thus, for $\Delta = 1$, which cannot actually be attained
in AdS/CFT, the hadron has a pointlike form factor.  For $\Delta >
1$ there is a suppression from the need to transfer momentum to all
partons~\cite{exclusive}.  For bulk fields with spin, the matrix
element contains an additional factor of $q^{s_A + s_B}$
and the effective number of partons is given by the twist $\tau =
\Delta - s$ \cite{hard}.

The crossover from the hard behavior occurs when $q Y^* \sim 1$, so
the effective size of hadrons is
\be
\< X^2 \> \sim Y^{*2}\ .
\ee
This corresponds to the confinement scale as measured by the masses
of the lightest hadrons, coming from supergravity modes.  As usual in
AdS/CFT duality, the length scale set by the flux tube tension is
smaller by a factor of $(gN)^{1/4}$, being given by the redshift of
the fundamental string scale,
\be
\alpha'^{1/2}_{\rm gauge\, theory} \sim \frac{Y^{*}}{R}
{\alpha'^{1/2}}
\sim (gN)^{-1/4} Y^{*}\ .
\ee
Thus the size is determined by the holographic spreading and not by
the internal wavefunction of the string, except for very highly
excited string states.  The detailed form of the form factor at small
$q$ can be obtained from the supergravity dual, but it depends on the
details of the conformal symmetry breaking.

These same methods can be applied to obtain other amplitudes such as
that for deep inelastic scattering, which is expressed in terms of 
two-current matrix elements~\cite{dis}.

\setcounter{equation}{0}
\section{Discussion }

The original attempt to describe hadrons as idealized strings
was frustrated by the infinite zero point oscillations in the size
of strings. Early ideas for modifying string theory such as
replacing the idealized strings by fat  flux tubes or as
collections of partons which approximate strings fit well with
QCD but seemed to preclude an idealized mathematical string
description.

More recent evidence from AdS/CFT type dualities suggests that
idealized
string theory in higher dimensions may provide an exact
description of the 't Hooft limit of QCD-like theories. In this
paper we have argued that an ideal bulk string theory in five
dimensions is fully compatible with a fat non-ideal string in four
dimensions.

The fifth dimension can be divided into two regions.  The ``wall"
region near  $Y=Y^*$ corresponds to
the confinement scale $\Lambda$. If we ignore high frequency
fluctuations,
the string spends most of its time in this region. The usual UV/IR
spreading gives the string a thickness of order $\Lambda$. High
frequency fluctuations of small sections of string can occur
which cause it to  fluctuate toward $Y=0$, the region
corresponding to short distance behavior in space-time.
These fluctuations  will control the large momentum
behavior of form factors as well as deep inelastic matrix
elements. Such fluctuations should give the string a parton-like makeup.
We have also seen that these fluctuations stiffen the effective
string tension so much that the infinite zero point size is
eliminated.

Vertex operators of the form (3.2) define matrix elements of local
currents such as the energy momentum tensor. Similar vertices can
be constructed for vector currents
of $R$-charge. Products of such vertices can be studied to uncover
the behavior of deep inelastic structure functions in the
confining deformations of conformal theories.

Our work and ref.~\cite{hard} show that even at large 't Hooft parameter
the short-distance behavior of gauge theories has much in common with
QCD.  However, to make contact with real QCD we need a better
understanding of the strongly coupled world-sheet theory at small $R$.
Even if we truncate to the single field $Y$ the theory~(\ref{gfact}) is
nontrivial, and it would be interesting to understand its behavior at
small $R$.

\section{Acknowledgements }

We would like to thank Juan Maldacena, Mimi Schwarz, and Matt
Strassler for helpful discussions.  This work was supported by NSF
grants PHY98-70115, PHY99-07949 and PHY00-98395.

\end{document}